\documentclass[prd,twocolumn,amsmath,amssymb,floatfix,superscriptaddress,notitlepage,nofootinbib,preprintnumbers]{revtex4-1}
\usepackage{amsfonts,amssymb,amsmath,graphicx,color,bm}
\definecolor{ultramarine}{rgb}{0.07, 0.04, 0.56}
\definecolor{cadmiumgreen}{rgb}{0.0, 0.42, 0.24}
\definecolor{indigo(dye)}{rgb}{0.0, 0.25, 0.42}
\usepackage[linktocpage=true]{hyperref}
\hypersetup{
colorlinks=true,
citecolor=ultramarine,
linkcolor=cadmiumgreen,
urlcolor=indigo(dye),
}

\newcommand{\para}[1]{\par\vspace{2mm}\noindent{\bf {#1}} --- }
\newcommand{\f}[2]{\frac{#1}{#2}}

\newcommand{\be}{\begin{equation}}  
\newcommand{\ee}{\end{equation}}
\newcommand{\bem}{\begin{pmatrix}}
\newcommand{\eem}{\end{pmatrix}}

\newcommand{\1}{\boldsymbol{1}}

\newcommand{\N}{\mathcal{N}}
\newcommand{\I}{\mathcal{I}}

\begin{document}

\title{Ghost-free theory with third-order time derivatives}

\author{Hayato Motohashi}
\affiliation{Center for Gravitational Physics, Yukawa Institute for Theoretical Physics, Kyoto University, Kyoto 606-8502, Japan}

\author{Teruaki Suyama}
\thanks{Present address: 
Department of Physics, Tokyo Institute of Technology,
2-12-1 Ookayama, Meguro-ku, Tokyo 152-8551, Japan}
\affiliation{Research Center for the Early Universe (RESCEU),
Graduate School of Science,
The University of Tokyo, Tokyo 113-0033, Japan}

\author{Masahide Yamaguchi}
\affiliation{Department of Physics, Tokyo Institute of Technology,
2-12-1 Ookayama, Meguro-ku, Tokyo 152-8551, Japan}

\begin{abstract}
As the first step to extend our understanding of higher-derivative
theories, within the framework of analytic mechanics of
point particles, we construct a ghost-free theory involving third-order time
derivatives in Lagrangian.  
While eliminating linear momentum terms in the Hamiltonian is necessary and sufficient to kill the ghosts associated with higher derivatives for Lagrangian with at most second-order derivatives,
we find that this is necessary but not sufficient for the Lagrangian with higher than second-order derivatives.
We clarify a set of ghost-free conditions under which we show that 
the Hamiltonian is bounded, and that equations of motion are reducible into a second-order system.
\end{abstract}

\preprint{YITP-18-31}

\maketitle

\para{Introduction}
In the last decade we have undergone a great progress of understanding 
ghost-free theories involving at most second-order time derivatives in Lagrangian~\cite{note1}.
The construction of higher-derivative theories requires special care 
to eliminate Ostrogradsky ghosts~\cite{Ostrogradsky:1850fid}.
While the Ostrogradsky theorem considers only ghost degrees of freedom (DOFs) 
associated with the highest-order derivatives, 
one needs to remove all the ghost DOFs associated with 
higher-order derivatives~\cite{Motohashi:2014opa}. 
All the proposed theories contain at most second-order time derivatives in their Lagrangian, 
for which all the Ostrogradsky ghosts can be evaded in a systematic way by 
imposing degeneracy condition of Lagrangian~\cite{Langlois:2015cwa,Motohashi:2016ftl} (see also Ref.~\cite{Klein:2016aiq}). 
The degeneracy condition applies for construction of various types of models~\cite{Crisostomi:2016czh,Kimura:2016rzw,BenAchour:2016fzp,Crisostomi:2017aim,Crisostomi:2017ugk,Kimura:2017gcy}.
In the picture of the Hamiltonian formulation,
the Ostrogradsky ghosts manifest themselves in the Hamiltonian as terms linear in the canonical momenta.
Eliminating linear momentum terms by means of
primary/secondary constraints has been thought to be
a principle to remove the ghosts associated with the higher derivative terms,
which actually works for the Lagrangian containing up to second-order derivatives.

In this paper, we perform the Dirac analysis
for a class of theories beyond the second-order time derivatives, 
and demonstrate how to obtain ghost-free theory involving third-order time derivatives in Lagrangian.
Contrary to the case with at most the second-derivatives in the Lagrangian, 
we explicitly confirm that elimination of the linear terms in momenta is necessary but not sufficient any more
to kill the ghosts for the Lagrangian containing up to third-order derivatives and that
ghosts are still hidden in the canonical coordinates corresponding to the higher time-derivatives.
Thus, in addition to the constraints used to remove the linear terms in momenta,
further constraints are required to obtain the Lagrangian which is completely free from the ghosts associated with higher-order derivatives.
We explicitly provide a set of conditions, under which we confirm that the Hamiltonian is bounded, and that equations of motion are reducible into a second-order system.

\para{Quadratic Model}
We consider a quadratic Lagrangian involving at most third-order time derivatives.
Lagrangian solely consisting of quadratic terms in the dynamical
variables, regardless of whether it contains higher derivative terms or not, 
is always enlightening and gives us a lot of insights \cite{note2}.

We consider the quadratic model defined by
\begin{align} 
L &= \f{a_{nm}}{2} \dddot\psi^n \dddot\psi^m + \f{b_{nm}}{2} \ddot\psi^n \ddot\psi^m + \f{c_{nm}}{2} \dot\psi^n \dot\psi^m  
\notag\\ 
&~~~
+ \f{d_{nm}}{2} \psi^n \psi^m + e_{nm} \dddot\psi^n \ddot\psi^m 
+ f_{nm} \ddot\psi^n \dot\psi^m  
\notag\\
&~~~ 
+ \f{A_{ij}}{2} \dot q^i \dot q^j + \f{B_{ij}}{2} q^i q^j + C_{ij} \dot q^i q^j
+ \alpha_{ni} \dddot\psi^n \dot q^i, \label{quad-lag}
\end{align}
where $q^i=q^i(t)$, $\psi^n=\psi^n(t)$, and $m,~n=1,\cdots, \N$, and $i,~j=1,\cdots,\I$.
We assume that $a_{nm}$, $b_{nm}$, $c_{nm}$, $d_{nm}$, $A_{ij}$, $B_{ij}$ are symmetric and
$e_{nm}$, $f_{nm}$, $C_{ij}$ are antisymmetric without loss of generality, 
and all the coefficients are nonzero constant for simplicity \cite{note3}.  
We also assume that there is no degeneracy in the $q^i$-sector,
$\det A_{ij} \neq 0$,
and denote its inverse matrix by $A^{ij}$. 
One may note that the coupling between $\psi^n$-system and $q^i$-system 
is governed by $\alpha_{ni}$ term.
As is explicitly shown in Refs.~\cite{Langlois:2015cwa,Motohashi:2016ftl}, 
a coupling between systems with different orders of derivatives leads to
a nontrivial degeneracy structure of the kinetic matrix, 
based on which various degenerate higher-order theories have been constructed~\cite{Crisostomi:2016czh,Kimura:2016rzw,BenAchour:2016fzp,Crisostomi:2017aim,Crisostomi:2017ugk,Kimura:2017gcy}. 
Below we shall see $\alpha_{ni}$ play a crucial role.

\para{Linear Momentum Terms}
To define conjugate momenta in the standard manner for the Hamiltonian analysis, 
we rewrite the Lagrangian~\eqref{quad-lag} 
by introducing auxiliary variables $R^n, Q^n$ and Lagrange multipliers $\xi_n, \lambda_n$
as
\begin{equation} 
L_{\rm eq} = 
L ({\dot Q},Q,R,\psi,{\dot q},q)+ \xi_n (\dot\psi^n - R^n) + \lambda_n (\dot R^n - Q^n). \label{Leq}
\end{equation}
The conjugate momenta 
for $(Q^n, R^n, \psi^n, q^i, \lambda_n, \xi_n)$ 
are then respectively given by
\begin{align} \label{canoqm}
P_{Q^n} &= a_{nm} \dot Q^m + \alpha_{ni} \dot q^i + e_{nm} Q^m, ~
P_{R^n}= \lambda_n, ~
\pi_{\psi^n}= \xi_n, \notag\\
p_i &= \alpha_{ni} \dot Q^n + A_{ij} \dot q^j + C_{ij} q^j, ~~
\rho_{\lambda_n}=0, ~~
\rho_{\xi_n}=0.
\end{align}

The last four equations yield $4\N$ primary constraints 
\begin{align} \label{Phis}
&\Phi_n\equiv P_{R^n}-\lambda_n \approx 0, \qquad
\Phi_{\N+n}\equiv \pi_{\psi^n} - \xi_n \approx 0, \notag\\
&\bar\Phi_{n}\equiv \rho_{\lambda_n} \approx 0, \qquad\qquad~~
\bar\Phi_{\N+n}\equiv \rho_{\xi_n} \approx 0.
\end{align}
Let us first assume that 
\be \label{nondeg} \det (a_{nm}-\alpha_{ni} A^{ij} \alpha_{mj}) \neq 0, \ee
which implies that ${\dot Q}^n$ and ${\dot q}^i$ can be solved as a function of ($P_{Q^n},p_i, Q^n,q^i$). 
In this case, \eqref{Phis} exhausts all the primary constraints and 
the Hamiltonian is given by
$H = H_0 + P_{R^n} Q^n + \pi_{\psi^n} R^n$, 
where 
$H_0$ is directly constructed from $L$ 
and, importantly, contains neither $P_{R^n}$ nor $\pi_{\psi^n}$.

From our understanding of theories involving at most second time derivatives in Lagrangian, 
ghost DOFs associated with higher derivatives always manifest themselves as 
terms linear in conjugate momenta in the Hamiltonian, 
and lead to the Hamiltonian neither bounded from below nor above. 
In the case of $L(\ddot\phi^a,\dot\phi^a,\phi^a;\dot q^i,q^i)$ considered in 
Ref.~\cite{Motohashi:2016ftl}, if we assume nondegeneracy corresponding to \eqref{nondeg}, 
the Hamiltonian 
depends linearly on the momentum conjugate to $\phi^a$.

Since we allow the third time derivatives in the Lagrangian \eqref{quad-lag}, 
the Hamiltonian $H$
has two linear momentum terms for 
$P_{R^n}$ and $\pi_{\psi^n}$ 
conjugate to $\dot\psi^n$ and $\psi^n$, respectively.
This exhibits the fact that the third derivatives introduce additional ghost DOFs, 
and it is natural to expect that a longer chain of the constraint algorithm should be required to remove all the ghost DOFs.
The linear dependences on momenta need to be removed by relating them to other variables by constraints.

If we assume \eqref{nondeg}, there is no more primary constraint other than \eqref{Phis}.
The evolution of the canonical variables is
generated by the total Hamiltonian 
$H_T = H + \mu_\alpha \Phi_\alpha + \bar\mu_\alpha \bar\Phi_\alpha$, where $\alpha=1,\cdots,2\N$ and
$\mu_\alpha$, ${\bar \mu}_\alpha$ are Lagrange multipliers.  
It can be easily verified that $\mu_\alpha$, ${\bar \mu}_\alpha$ are fixed by 
consistency conditions of the primary constraints~\eqref{Phis}
and Dirac algorithm terminates here without any secondary constraints.
Since the primary constraints do not relate $P_{R^n}$ nor $\pi_{\psi^n}$ with other variables in $H_0$, 
$H$ is neither bounded from below
nor above and suffers from the Ostrogradsky ghosts.
This is a natural consequence of the Ostrogradsky theorem.

\para{Eliminating Linear Momentum Terms}
A way out to evade the Ostrogradsky ghosts is to violate 
\eqref{nondeg}. 
We require all the eigenvalues of the matrix $a_{nm}-\alpha_{ni} A^{ij} \alpha_{mj}$ are vanishing 
because otherwise there remain ghost DOFs corresponding to nonvanishing eigenvalues.
Thus, we impose
\be \label{DC1qm} a_{nm}-\alpha_{ni} A^{ij} \alpha_{mj}=0, \ee
which we call the first degeneracy condition (DC1).
When the Lagrangian satisfies DC1, the first two equation of \eqref{canoqm} 
yield additional $\N$ primary constraints 
\be \label{Psis} \Psi_n\equiv P_{Q^n}-e_{nm} Q^m -A^{ij}\alpha_{nj} \tilde p_i \approx 0 , \ee
where ${\tilde p}_i \equiv p_i-C_{ik}q^k$, 
with which $H_0$ simplifies as
$H_0 = \f{1}{2} A^{ij} \tilde p_i \tilde p_j 
- \f{1}{2} b_{nm}Q^nQ^m  
- \f{1}{2} c_{nm}R^nR^m
- \f{1}{2} d_{nm}\psi^n\psi^m 
- f_{nm}Q^nR^m - \f{1}{2} B_{ij} q^iq^j$,
and 
the total Hamiltonian
is modified as 
$H_T = H + \mu_\alpha \Phi_\alpha + \bar\mu_\alpha \bar\Phi_\alpha + \nu_n \Psi_n$.
However note that we still need further constraints as
\eqref{Psis} constrains neither $P_{R^n}$ nor $\pi_{\psi^n}$.
The consistency conditions $\dot\Phi_\alpha \approx 0$, $\dot{\bar\Phi}_\alpha \approx 0$ 
respectively fix $\bar\mu_\alpha,\mu_\alpha$, whereas $\dot\Psi_n \approx 0$ yields
\be \{\Psi_n,H\} + \nu_m \{\Psi_n,\Psi_m \} \approx 0 . \label{d-Psi}  \ee
If $\det\{ \Psi_n,\Psi_m \}\neq 0$, $\nu_m$ is uniquely fixed and the Dirac algorithm terminates here, 
and the Hamiltonian is left with the linear momentum terms.
Therefore, the avoidance of the Ostrogradsky theorem by 
imposing DC1 is 
not sufficient to eliminate all the ghost DOFs.
Such a case was first pointed out in \cite{Motohashi:2014opa} for general Lagrangian involving arbitrary higher-order derivatives.
The reason is that the Ostrogradsky theorem only considers the ghost DOFs associated with the highest derivatives. 
In the present case, the linear terms in $P_{R^n}$ and $\pi_{\psi^n}$ in the Hamiltonian are ghosts associated with non-highest but higher-order derivatives.

To eliminate the ghost DOFs, we require that 
the consistency conditions~\eqref{d-Psi} do not determine $\nu_n$.
In parallel to the DC1, we 
impose DC2 as
\be \label{DC2qm} \{ \Psi_n,\Psi_m \}=-2 [e_{nm} + \alpha_{ni} {(A^{-1}CA^{-1})}^{ij} \alpha_{mj}] =0, \ee
with which \eqref{d-Psi} yields the secondary constraints 
as
\begin{align} \label{C2qm} 
&0 \approx \Upsilon_n \equiv - \{\Psi_n,H \} = P_{R^n} - b_{nm}Q^m - f_{nm}R^m 
\notag\\
&
- 2 \alpha_{ni} {({\bar C} A^{-1})}^{ij} \tilde p_j +\alpha_{ni} {\bar B}^i_{~j} q^j,  
\end{align}
where 
${\bar B} \equiv A^{-1}B$ and ${\bar C} \equiv A^{-1}C$. 
These constraints fix $P_{R^n}$ in terms of
other variables and eliminate the ghosts associated with
the linear terms in $P_{R^n}$ in $H$.

This is parallel to the case of $L(\ddot\phi^a,\dot\phi^a,\phi^a;\dot q^i,q^i)$ 
considered in Ref.~\cite{Motohashi:2016ftl}, 
where the linear dependencies on the momenta conjugate to $\phi^a$ are removed by 
the secondary constraints that arise by DC2.
In that case all the linear dependencies are removed and 
the program of removing the ghosts is achieved with DC1 and DC2.

In the present case, DC1 and DC2 are not sufficient 
as there still remain the linear terms on $\pi_{\psi^n}$. 
In order to eliminate them, consistent time evolution for $\Upsilon_n \approx 0$, 
\be \{\Upsilon_n,H\} + \nu_m \{\Upsilon_n,\Psi_m \} \approx 0 , \label{d-Up} \ee
must yield tertiary constraints otherwise the Dirac algorithm terminates here.  
We thus impose DC3 as
\be \label{DC3qm} \{ \Upsilon_n, \Psi_m \} =
-b_{nm}-\alpha_{ni} [(4{\bar C}^2+{\bar B}) A^{-1} ]^{ij} \alpha_{mj}=0, \ee
with which
\eqref{d-Up} yields the tertiary constraints
\begin{align} \label{C3qm}
&0 \approx \Lambda_n\equiv -\{ \Upsilon_n, H \} = \pi_{\psi^n} + 2f_{nm}Q^m - c_{nm}R^m 
\notag\\
&
+2 \alpha_{ni} {({\bar C} {\bar B} )}^i_{~j} q^j 
-\alpha_{ni} [(4{\bar C}^2+{\bar B}) A^{-1} ]^{ij} \tilde p_j ,
\end{align}
which fix $\pi_{\psi^n}$ in terms of other variables.
Here, note that $\{\tilde p_i, \tilde p_j\}=-2C_{ij}$.
Thus, all the terms linear in $P_{R^n}$ and $\pi_{\psi^n}$ in $H$ are fixed by imposing DC1 -- DC3,
and it is natural to expect that we completed the reduction of the theory~\eqref{Leq} to a healthy one.

\para{Hidden Ghosts}
Now we encounter a highly nontrivial situation that never
happens in theories involving at most second-order time
derivatives in the Lagrangian:
the above expectation is not the case and the Hamiltonian still contains ghost DOFs.
To see this explicitly, let us suppose that the consistent time evolution of the tertiary constraints~\eqref{C3qm}
\be \{\Lambda_n,H\} + \nu_m \{\Lambda_n,\Psi_m \} \approx 0 , \label{d-Lam} \ee
would just fix the Lagrange multipliers $\nu_n$ and we would finish the Dirac algorithm.  
Note that the variables $(Q^n, R^n,\psi^n,q^i,p_i)$ remain unconstrained.  
Even though we fixed all the terms linear in momenta in the Hamiltonian, 
since $Q^n=\ddot\psi^n$ is not constrained, 
we expect this system still have ghost DOFs.
Indeed, by plugging \eqref{C2qm}, \eqref{DC3qm}, \eqref{C3qm} into $H$, 
we can erase $P_{R^n}$, $b_{nm}$, $\pi_{\psi^n}$ and arrive at 
\begin{align} \label{quad-ham}
H&=\frac{1}{2} A^{ij} {\bar p}_i {\bar p}_j-\frac{1}{2} B_{ij} {\bar q}^i {\bar q}^j+\frac{1}{2} c_{nm}R^n R^m
-\frac{1}{2}d_{nm} \psi^n \psi^m 
\notag\\
&~~~ 
+\alpha_{ni} [(4{\bar C}^2+{\bar B})A^{-1}]^{ij} {\bar p}_j R^n 
-2 \alpha_{ni} {({\bar C} {\bar B})}^i_{~j} {\bar q}^j R^n  \notag\\
&~~~
-2 [ f_{nm}+4 \alpha_{ni} {({\bar C}^3 A^{-1})}^{ij} \alpha_{mj} ] Q^m R^n, 
\end{align}
where ${\bar p}_i \equiv \tilde p_i+2 \alpha_{nk} {\bar C}^k_{~i}Q^n$ and 
${\bar q}^i \equiv q^i+\alpha_{nk} A^{ki} Q^n$.
Remarkably, 
all the quadratic terms in $Q^n$ have been absorbed in those of ${\bar p}_i$
and ${\bar q}^i$, and $Q^n$ appears in \eqref{quad-ham} only linearly,
making $H$ unbounded.

The important lesson here is that, 
beyond second-order derivatives,
ghost DOFs associated with higher derivatives do not 
precisely correspond to the linear dependencies on momenta in the Hamiltonian.  
Eliminating the linear momentum terms by the constraints 
is always necessary condition to kill the Ostrogradsky ghosts,
but is not sufficient condition for Lagrangians containing higher-than-second-order derivatives. 
In the latter case, after the linear momentum terms are eliminated,
the ghosts still lurk in the Hamiltonian in a very nontrivial manner.
For the present model, we could prove the existence of the hidden ghosts
in \eqref{quad-ham} explicitly.

\para{Eliminating Hidden Ghosts}
To tame the terms linear in $Q^n$ in \eqref{quad-ham}, impose DC4 as
\be \label{DC4qm} \{ \Lambda_n, \Psi_m \}=2 ( f_{nm}- \alpha_{ni} 
M^{ij} \alpha_{mj} ) =0. 
\ee 
where
$M\equiv (4{\bar C}^3 +{\bar B}{\bar C}+{\bar C}{\bar B})A^{-1}$.
We then obtain the quaternary constraints from \eqref{d-Lam} as
\begin{align} \label{C4qm}
&0\approx \Omega_n \equiv -\{ \Lambda_n, H \} =c_{nm}Q^m-d_{nm}\psi^m
\notag\\
&
-2 \alpha_{ni} M^{ij} \tilde p_j 
+\alpha_{ni} {(4{\bar C}^2 {\bar B}+{\bar B}^2)}^i_{~j} q^j,
\end{align}
which, assuming $\det c_{nm}\neq 0$, 
fix $Q^n$ in terms of other variables, and hence the Hamiltonian~\eqref{quad-ham} is bounded.  
Unconstrained variables are now $(R^n, \psi^n, q^i, p_i)$,
and there are no longer unconstrained variables
corresponding to higher derivatives.

While in general the consistency evolution of 
the quaternary constraints~\eqref{C4qm} could generate further
constraints, such constraints inevitably kill the healthy DOFs.  
Since our purpose is to kill only the ghost DOFs
associated with higher derivatives, we require that the Dirac
algorithm terminates here by imposing $\det Z_{nm} \neq 0$,
where 
$Z_{nm}\equiv \{ \Omega_n, \Psi_m \}=c_{nm}-\alpha_{ni} N^{ij} \alpha_{mj}$
with 
$N\equiv 4\bar C M + (4 \bar B \bar C^2 + \bar B^2)A^{-1}$.
Since the matrix $c_{nm}$ has nothing to do with $(\alpha_{ni}, A, B, C)$,
the requirement $\det Z_{nm} \neq 0$ can be met in general.
In this case $\nu_n$ is determined by
\be \nu_n =Z^{nm}( d_{mm'}{\dot \psi}^{m'}-\alpha_{mi}N^{ij} {\tilde p}_j+2 \alpha_{mi}M^{ij} B_{jk}q^k ). \label{quad-eq-nu} \ee

\para{Number of DOFs}
The Dirac matrix is given by
\be
\begin{array}{l|cccccc}
  & \Phi_\beta & \bar\Phi_\beta & \Psi_m & \Omega_m & \Upsilon_m & \Lambda_m  \\ \hline
  \Phi_\alpha     & 0 & -\1 & * & * & * & * \\
  \bar\Phi_\alpha & \1 & 0 & 0 & 0 & 0  & 0 \\
  \Psi_n          & * & 0 & 0 & -Z_{mn} & 0 &0    \\
  \Omega_n    & * & 0 & Z_{nm} & * & *  & * \\
  \Upsilon_n      & * & 0 & 0 & * & 0 &Z_{mn}  \\
  \Lambda_n      & * & 0 & 0 & * & -Z_{nm} &*  \\
\end{array} 
\ee
where the matrix elements marked by $*$ are irrelevant to the determinant of the Dirac matrix, and we have used the identities 
$\{ \Upsilon_n,\Upsilon_m \}=0$ and $\{ \Upsilon_n,\Lambda_m \}=Z_{mn}$, 
which can be shown by plugging $\Upsilon_n\equiv\{H,\Psi_n \}$, Jacobi identity, DC3~\eqref{DC3qm}, and DC4~\eqref{DC4qm}. 
Now it is clear that the determinant of the Dirac matrix is given by
$(\det Z)^4$, which is nonvanishing.
Thus, all the $8\N$ constraints are second class. 
The number of DOFs is then given by
$( 10 \N+2\I-8\N )/2=\N+\I$,
which precisely coincides with the number of variables.

In summary,
the above Hamiltonian analysis demonstrates that 
fixing the linear dependencies on momenta $(P_{R^n},\pi_{\psi^n})$ is not sufficient and 
the Dirac algorithm must continue until $(P_{Q^n},Q^m)$ are fixed.
They are fixed by the constraints that arise by imposing DC1 -- DC4.
Consequently, we obtain the Lagrangian with unconstrained variables $(R^n,\psi^n,q^i,p_i)$, 
the bounded Hamiltonian, and the healthy number of DOFs.
As expected, the nonzero coupling $\alpha_{ni}$ play a crucial role.
If $\alpha_{ni}=0$ and hence $\psi^n$-system and $q^i$-system are decoupled, 
DC1 -- DC4 are nothing but requiring coefficients for higher-derivative terms are identically vanishing.
It is thus precisely the nonzero coupling 
that leads to a nontrivial degeneracy structure of the theory.

\para{Hamiltonian Equations}
From the number of DOFs, we naively
expect that the Hamilton equations for the unconstrained variables $(
R^n, \psi^n, q^i, p_i )$ can be reduced to a closed set of second-order
differential equations for $\psi^n$ and $q^i$.  
Below we show that this is indeed the case.  
The Hamilton equations are given by
\begin{align}
&\dot R^n=Q^n, \quad 
\dot \psi^n =R^n, \quad
\dot q^i=A^{ij} ({\tilde p}_j-\nu_n \alpha_{nj}), \label{quad-H-q} \\
&\dot{\tilde p}_i= B_{ij} (q^j+\alpha_{nk}A^{kj}Q^n ) + 2 \alpha_{nj} ({\bar C}\bar B)^j_{~i} R^n. 
\label{quad-H-p}
\end{align}
By using \eqref{C4qm}, \eqref{quad-eq-nu}, the third equation of \eqref{quad-H-q} to the right-hand side of 
the first equation of \eqref{quad-H-q},
we obtain  
\be \label{ddpsi} \ddot \psi^n = \text{(lower derivatives)}. \ee
Likewise, from \eqref{quad-H-p} we obtain 
\be \label{quad-L-q} J_i{}^j A_{jk} \ddot q^k=\text{(lower derivatives)}, \ee
where $J_i{}^j \equiv \delta_i{}^j-\alpha_{ni} c^{nm}\alpha_{mk} N^{kj}$.
Since $A_{jk}$ is regular by definition, if $\det J \neq 0$, 
we can multiply an inverse matrix to express $\ddot q^i$ in terms of lower derivatives.
Indeed, through proof by contradiction, we can show $\det J \neq 0$ as follows.
If we assume $\det J=0$, then there exists a nonzero vector $V^i$ that is mapped to a null vector by the operation of $J_i{}^j$, i.e.\ $V^i J_i{}^j=0$, from which we obtain  
$0=V^i J_i{}^j\alpha_{nj}=V^i \alpha_{m'i} c^{m'm} Z_{mn}$, 
where we used an identity $J_i{}^j \alpha_{nj}=\alpha_{m' i}c^{m'm} Z_{mn}$.
Clearly, it contradicts with the regularity of $c^{mm'} Z_{m'n}$ by 
$\det c_{nm}\neq 0$ and $\det Z_{nm} \neq 0$, and hence $\det J \neq 0$ holds. 
Thus, from \eqref{quad-L-q} we obtain 
\be \label{ddq} \ddot q^i = \text{(lower derivatives)}. \ee
Here we completed the reduction of the original Hamilton equations to a closed set of second-order differential equations \eqref{ddpsi} and \eqref{ddq}.
It is also worthwhile to note that the condition $\det Z_{nm} \neq 0$ for the termination of the Dirac algorithm precisely terminates the reduction of the equations of motion.

\para{Euler-Lagrange Equations}
Now, it is intriguing to investigate Euler-Lagrange (EL) equations 
when the phase space is reduced as shown in the Hamiltonian analysis.  
To this end, we go back to the original Lagrangian \eqref{quad-lag}.
While the EL equations for $\psi^n$ and $q^i$ a priori 
contain up to sixth-order time derivatives,
below we show that 
they can be reduced to the second-order system expressed by \eqref{ddpsi} and \eqref{ddq}
so long as the four DCs, \eqref{DC1qm}, \eqref{DC2qm}, \eqref{DC3qm}, \eqref{DC4qm}, 
as well as the condition 
$\det Z_{nm} \neq 0$ are imposed.

First we rewrite the EL equation for $q^i$ as
\be {\ddot q}^i=A^{ij} ( -\alpha_{nj} \psi^{n(4)}-2 C_{jk} {\dot q}^k+B_{jk}q^k ). \label{quad-eq-ddotq} \ee
Plugging it into the EL equation for $\psi^n$, we note that 
the coefficients for $\psi^{n(6)}$, $\psi^{n(5)}$, $\psi^{n(4)}$ vanish as they precisely match the DC1~\eqref{DC1qm}, DC2~\eqref{DC2qm}, DC3~\eqref{DC3qm}. 
By erasing $f_{nm}$ 
by plugging DC4~\eqref{DC4qm}, we obtain
\begin{align}
&c_{nm} {\ddot \psi}^m-d_{nm} \psi^m+\alpha_{ni} {(4{\bar C}^2 {\bar B}+{\bar B}^2 )}^i_{~j} q^j
\notag\\
&
=2 \alpha_{ni} {(MA )}^i_{~j} ( {\dot q}^j+A^{jk} \alpha_{mk} {\dddot \psi}^m). \label{quad-EL-3psi}
\end{align}
While \eqref{quad-eq-ddotq} and \eqref{quad-EL-3psi} still contain
$\psi^{n(4)}$ and $\dddot\psi^n$, they can be removed as follows.
Taking a time derivative of \eqref{quad-EL-3psi}, we can use \eqref{quad-eq-ddotq} to the right-hand side and obtain  
\be
c_{nm} {\dddot \psi}^m-d_{nm} {\dot \psi}^m + \alpha_{ni} (NA)^i_{~j} {\dot q}^j-2 \alpha_{ni} {(MB)}^i_{~j} q^j = 0 . \label{quad-EL-2q} 
\ee
Since the matrix $c_{nm}$ is assumed to be regular, \eqref{quad-EL-2q} enables us 
to express ${\dddot \psi}^n$ in terms of at most first derivatives.  
We can then replace ${\dddot \psi}^m$ in \eqref{quad-EL-3psi} 
to obtain \eqref{ddpsi}.
Taking a time derivative of \eqref{quad-EL-2q}, and plugging \eqref{ddpsi} to erase $\ddot\psi^n$, we obtain an equation $\psi^{n(4)}=-c^{nm}\alpha_{mi}(NA)^i_{~j}\ddot q^j +$(lower derivatives).
Plugging this equation to \eqref{quad-eq-ddotq}, we obtain \eqref{quad-L-q}.
As $\det J \neq 0$ holds by virtue of 
$\det c_{nm}\neq 0$ and $\det Z_{nm} \neq 0$, we obtain \eqref{ddq}.

\para{Reduction of Higher-Derivative Degenerate Lagrangian 
to an Equivalent Lower-Derivative
Nondegenerate Lagrangian}
Finally, we can show that under DC1 -- DC4 the Lagrangian~\eqref{quad-lag} 
is equivalent to a Lagrangian involving at most first order derivatives. 
Indeed, using DC1 -- DC4 and a redefinition of variable
$r^i = q^i+\alpha_{nk} A^{ki} {\ddot \psi}^n + 2\alpha_{nk}(\bar C A^{-1})^{ki}\dot\psi^n$,
one can immediately see that the Lagrangian~\eqref{quad-lag} is reduced to a Lagrangian written up to the first-order derivative of $\psi^n$ and $r^i$.

One might think that it is possible to generalize this result to any degenerate Lagrangian that contains higher derivatives but has only healthy DOFs.
Below
we show that if all the constraints are second class this conjecture is true, 
namely, there always exist 
new variables defined by the 
transformation from the original ones 
such that the new Lagrangian in terms of the new variables is nondegenerate, contains at most the first-order time derivatives, and still describes the dynamics defined by the original Lagrangian.

Let us consider a degenerate Lagrangian with higher derivatives that contains only healthy DOFs.
In the language of the Hamiltonian analysis, this amounts to the presence of the constraints 
by which the Ostrogradsky modes are eliminated.
Let us write the degenerate Lagrangian with higher derivatives as $L(q^i,{\dot q}^i,{\ddot q}^i,\cdots)$,
where $\cdots$ represents higher-order time derivatives of $q^i$. 
Without loss of generality, by means of Lagrange multipliers,
we can rewrite this Lagrangian into new one that contains at most first-order time derivatives
as $L({\tilde q}^I,\dot {\tilde q}^I)$.
In order to move to the Hamiltonian formalism, we introduce the conjugate momenta ${\tilde p}_I$
by the standard definition.
By our assumption that this Hamiltonian is free from the Ostrogradsky ghosts,
there are a chain of primary and secondary (and possibly more) constraints 
$\Phi_A ({\tilde q}^I,{\tilde p}_I) \approx 0$.
We assume all the constraints are second class.
Then, according to 
Ref.~\cite{Maskawa:1976hw},
it is always possible to
perform a canonical transformation such that the new canonical variables are
decomposed into two sets $(Q^a,P_a)$ and $({\tilde Q}^m,{\tilde P}_m)$, and the
constraints are represented as ${\tilde Q}^m \approx 0,~{\tilde P}_m \approx 0$.
Thus, time evolution of the unconstrained variables $(Q^a,P_a)$ is governed by the Hamiltonian
$H(Q^a,P_a)$ that contains only unconstrained variables.
In other words, trajectory in the phase space is confined to the subspace spanned
by $(Q^a,P_a)$ and 
is governed by the Hamiltonian equation ${\dot Q}^a=\frac{\partial H}{\partial P_a}$, ${\dot P}_a=-\frac{\partial H}{\partial Q^a}$
as if no constraints were imposed in the system.
Now, since $(Q^a,P_a)$ are unconstrained variables, it is possible to solve the
first equation above for $P_a$ as $P_a=P_a (Q^b,{\dot Q}^b)$.
Then, using this relation, we can construct the corresponding Lagrangian in the standard way as
$L'(Q^a,{\dot Q}^a)=P_a {\dot Q}^a-H(Q^a,P_a)$.
This Lagrangian is nondegenerate and contains up to first-order time derivatives of $Q^a$.

Finally, even if a theory contains first-class constraints,
the above result applies if 
by gauge fixing the theory can be transformed to the one with only second-class constraints.

\para{Conclusion and Discussion}
It has been shown in the previous works~\cite{Motohashi:2014opa,Langlois:2015cwa,Motohashi:2016ftl}
that the condition for the absence of Ostrogradsky ghosts for theories
with second-order derivatives is to eliminate linear dependence of
canonical momenta in the Hamiltonian by imposing a certain set of
ghost-free conditions.  
In this paper, we found that eliminating linear dependence of
canonical momenta in the Hamiltonian is not sufficient for theories with
higher-than-second-order derivatives, and that canonical coordinates
corresponding to the higher time-derivatives also need to be removed
appropriately.  We stress that this feature shows up only in theories
with higher-than-second order time derivatives in the Lagrangian. 
We have constructed an illuminating example of
ghost-free higher-order theories by imposing a
certain degenerate conditions, under which we showed that the Hamiltonian
equations and EL equations are reducible to a second-order
system. 
Furthermore, we showed that for general degenerate theory with higher derivatives
but possessing only healthy DOFs, 
if all the constraints are second class, it is always possible
to define new variables 
such that the new Lagrangian in terms of the new variables is nondegenerate, contains at most the first-order time derivatives, and still describes the dynamics defined by the original Lagrangian.

Throughout the present paper, 
we restricted ourselves to 
the quadratic model involving at most third-order derivatives 
as it is the minimal system
beyond Lagrangian involving at most second-order derivatives
exhibiting the crucial difference.
However, we emphasize that the resultant criteria of ghost-free theory are quite reasonable 
and its physical meaning is very clear.
In the forthcoming paper~\cite{Motohashi:2018pxg}, we will generalize the present
analysis to more general system beyond quadratic model including not only a generic action with
up to third order derivatives but also that with arbitrary higher
derivatives.  It would be also
interesting to employ the proposed model to construct the
corresponding field theory and to investigate its phenomenology.

\vspace{1mm}

\para{Acknowledgments}
This work was supported in part by JSPS KEKENHI Grant Numbers,
JP17H06359 (H.M.), JP25287054 (M.Y.), JP26610062 (M.Y.).
JP17H06359 (T.S.), JP15H05888 (T.S.\ \& M.Y.), and JP15K17632 (T.S.).

\bibliography{ref-3deriv}

\begin{thebibliography}{40}
\bibitem [{not({\natexlab{c}})}]{note3}%
  \BibitemOpen
  \href@noop {} {} \bibinfo {note} {Of course, the
  Lagrangian~\eqref{quad-lag} is not the most general quadratic Lagrangian. For
  instance, one could include ${\ddot \psi}^n {\dot q}^i$. It is also possible
  to add other variables that appear in Lagrangian up to their second
  derivatives. However, since the nontrivial structure of the Hamiltonian
  analysis is already sharply visible in this simple setup, we focus on the
  model~\eqref{quad-lag}.}\BibitemShut {Stop}%
\end{thebibliography}

\end{document}